\documentclass[12pt, a4paper]{article}

\usepackage{graphicx}
\usepackage{amssymb}
\usepackage{xspace}
\usepackage{amsmath}
\usepackage[colorlinks=true, allcolors=blue]{hyperref} 
\usepackage{authblk}
\usepackage[numbers]{natbib}

\newcommand{\DT}{DT\@\xspace} %
\newcommand{\DTs}{DTs\@\xspace}  %

\providecommand{\keywords}
{
  \small	
  \textbf{Keywords} 
} 

\providecommand{\sep}
{
$\cdot$
}

\title{A Data Taxonomy Towards the Applicability of the Digital Twin Conceptual Framework in Disaster Management}

\author[1]{Eva Brucherseifer}
\author[2]{Marco Marquard}
\author[3]{Martin Hellmann}
\author[2]{Andrea Tundis}

\affil[1]{\small{University of Applied Science Darmstadt, Schöfferstr. 8, 64295 Darmstadt, Germany}}
\affil[2]{Institute for the Protection of Terrestrial Infrastructures (PI), German Aerospace Center (DLR), Rathausallee 12, St. Augustin, 53757, Germany}
\affil[3]{University of Applied Sciences Cologne, Schaevenstraße 1B, 50676 Köln, Germany}

\date{July 2024}

\begin{document}

\maketitle

\begin{abstract}
The Digital Twin (DT) offers a novel approach to the management of critical infrastructures, including energy, water, traffic, public health, and communication systems, which are indispensable for the functioning of modern societies. The increasing complexity and interconnectedness of these infrastructures necessitate the development of robust disaster response and management strategies. During crises and disasters, data source availability for critical infrastructure may be severely constrained due to physical damage to communication networks, power outages, overwhelmed systems, sensor failure or intentional disruptions, hampering the ability to effectively monitor, manage, and respond to emergencies.

This research introduces a taxonomy and similarity function for comparing data sources based on their features and vulnerability to crisis events. This assessment enables the identification of similar, complementary, and alternative data sources and rapid adaptation when primary sources fail. The paper outlines a data source manager as an additional component for existing DT frameworks, specifically the data ingress and scenario mangement. A case study for traffic data sources in an urban scenario demonstrates the proposed methodology and its effectiveness.
This approach enhances the robustness and adaptability of DTs in disaster management applications, contributing to improved decision-making and response capabilities in critical situations.
\end{abstract}

\keywords Digital Twin \sep Critical Infrastructure \sep  Disaster Management \sep Data Taxonomy \sep Data Source Management \sep Data Classification 

\section{Introduction} \label{sec:intro}

In an era of increasing complexity and interconnectedness of urban systems and critical infrastructures, the need for advanced tools to manage and respond to disasters has never been more pressing. Digital Twins (DTs) have emerged as a promising technology for enhancing disaster preparedness and response capabilities. A Digital Twin is a dynamic virtual replica of a physical entity or system that maintains a two-way flow of information, enabling real-time communication and interaction between the physical object and its digital counterpart. This bidirectional link facilitates continuous synchronization, analysis, simulation, and optimization, providing valuable operational insights crucial for preemptive decision-making during crises.

While the potential of DTs in disaster management is significant, their effectiveness hinges on the availability and reliability of data inputs. In crisis situations, the failure of critical infrastructures or damage to sensor networks can severely impact data availability, potentially compromising the DT's ability to accurately represent and simulate real-world conditions. This vulnerability underscores the need for robust data management strategies that can adapt to changing circumstances and maintain the DT's functionality even in the face of data source disruptions.

This research aims to investigate on assessment methods of data sources. In this paper we propose a classification based approach resulting in the concept of a Data Source Management component for the DT. Such a component provides a replacement algorithm taking data features and vulnerability of data source into account. A case study on traffic data sources will be provided.

The significance of this work lies in its potential to bridge the gap between theoretical DT frameworks and their practical implementation in disaster management scenarios. By providing a structured approach to data source classification and management, this research aims to enhance the robustness and adaptability of Digital Twins in crisis situations. This, in turn, can lead to more effective disaster response and management practices, ultimately contributing to improved resilience of urban systems and critical infrastructures.

The remainder of the paper is organized as follows: Section \ref{sec:related_work} provides an overview of related work in data classification for crisis management and Digital Twin applications. Section \ref{sec:research_questions} outlines the methodological approach and research questions. Section \ref{sec:ds_classification} details the proposed data classification scheme and replacement mechanism. Section \ref{sec:dt_framework} discusses the application of the DT conceptual framework to disaster data, including the integration of the proposed data management approach. Section \ref{sec:ex} provides a case study for the proposed methods. Finally, Section \ref{sec:conclusion} concludes the paper with a summary of findings and directions for future research.

\section{Related Work}
\label{sec:related_work}

\subsection{Data for crisis management}

While crisis management relies on information about the crisis, its impact and development, the processing of the data plays an important role.
The used data should provide an overview of the situation and enable the crisis managers to draw the right conclusions.
With the availability of data from multiple sources nowadays the challenge is not only to gather data about the crisis but also to interpret it in the right way.
We have found multiple approaches in the literature to this challenge.

In publications \cite{Conges2023} and \cite{Essendorfer2020} approaches to present data in comprehensible way through visualization techniques are discussed.
The focus of \cite{Conges2023} is the usage of virtual reality to create a crisis management cell.
In this cell the data is presented in an immersive way to the crisis managers and domain experts can join the cell from remote places to support the decision making with their expertise.
An alternative approach using visual analytics in crisis management as well as a multidisciplinary approach combining visualization and data analysis, is discussed in \cite{Essendorfer2020}.
The approach aims to provide a deeper insight into heterogeneous data used in crisis management and thus enhance the decision making process.
Additionally the paper elaborates on the benefits of Coalition Shared Data (CSD), but only mentions the potential data analysis techniques marginally.

A recent research contribution \cite{Zheng2023} has adopted incoming DT data analysis to deal with real-time detection of cybertheats using anomaly detection operation by applying a multidimensional deconvolutional network equipped with an attention mechanism.
An ontology-based approach has been recently published in \cite{Bai2024}, where an information framework for tracking provenance and managing information has been embedded in knowledge graphs.
In \cite{Borges2023} a taxonomy has been built to better understanding topics, interrelations, challenges, gaps related to crisis information managment systems.

A broad overview about the potential of the analysis of spatio-temporal data with deep learning techniques is presented in \cite{Wang2022}.
The survey covers multiple domains where deep learning approaches demonstrated their potential.
These domains include relevant domains for crisis management like human mobility and crime analysis.
Even though the analysis of big data benefits from deep learning approaches there are open research questions which hinder the usage in crisis management.
Deep learning models are often considered as black boxes and thus make it hard to interpret and explain the results.
While the analysis of large data sets is a common application for deep learning models, the fusion of heterogeneous spatio-temporal data sets is still an active area of research.
The potential of spatial-temporal selection criteria are shown in \cite{Zoppi2018}, where information collected by a crisis management system from different sensors including human sensors are labelled by reference using the spatial and temporal information of the data sources.
The benefits of using an established standard for the data collection is discussed in \cite{Lopez2019}.
The usage of connectors and data converters enable the storage of data from sources with differing standards in common database and the provision of the information to varying systems used in crisis management.

The handling of big data during crisis management is being discussed in \cite{Shah2019}. The authors give a thorough overview of the overall architectural deployment of big data analytics- and Internet of Things-based disaster management environments through a reference model having dedicated layers, such as data generation, harvesting, communication, management and analytics, and applications.
In a similar approach \cite{Li2022} applies big data analysis on the massive data generated in the smart city. Deep learning algorithms are being used to analyse electricity performance. The authors claim increased efficiency and reduced data usage. However, deep learning requires prior availability of data sets. As each disaster event is different the approach fails in crisis management.
The authors in \cite{Nayak2023} apply deep learning to enhance IoT analytics and teaching. It is advertised to integrated deep learning into the IoT sensors.

Together with \cite{Palen2016} the above mentioned publications \cite{Zoppi2018} and \cite{Lopez2019} describe the additional challenges to gather information from social media or integrate it as additional source.
In \cite{Palen2016} it is shown that social media provides new insights in crisis scenarios.
As information is often distributed over multiple short messages and the context can only be understood by analyzing all related messages, even simple questions require mature analysis approaches to extract the relevant information.
Also the management of misinformation, noise and bias in social media requieres advanced forms of filtering and verification techniques \cite{Zoppi2018} to select only valid and trusted data.

The existing research contributions address numerous aspects of the data processing chain in crisis management, spanning from data acquisition to data analysis and decision support.  However, none of the contributions addressed the full data life cycle, including the issue of data source interruptions.

\subsection{Digital Twin Application in Crisis Management}
\label{sec:related_work:Framework}

The concept of digital twins has been adopted for the monitoring, analysis, control and optimization of systems throughout their life cycle with a focus on real-time interaction.
Virtual replication and twinning form the core concept for synchronizing the \DT with the real world section, which is highly dependent on availability of reliable data.
Implementation details include considerations for data acquisition, modeling, value creation, human-computer interfaces and platform implementation. As Figure \ref{fig:digital_twin_basic} depicts added value is created by tools that offer smart functionality as part of the \DT, predict future conditions and initiate smart measures.
This makes \DTs a technology suitable for use in crisis management and promises to improve the resilience of critical infrastructure services, highlighting its responsiveness, intelligent features and adaptability.
The following literature research displays a variety of proposed implementation approaches.

The term "Digital Twin" was first introduced in 2002 by Grieves \cite{Grieves2014}. Over the years, the concept has undergone intense research and development efforts as surveyed by Fuller \cite{Fuller2020}, Liu \cite{Liu2021}, and Jones \cite{Jones2020}. This sets the historical context and underscores the increasing significance of \DTs in various domains.

In \cite{Galera-Zarco2023}, the authors explored the  role of DT beyond enhancing internal processes, recognizing their potential to shape new service offerings. They highlighted the need to investigate DT's role in enabling services, prompting the creation of a taxonomy classifying these digital services. Through literature review and use cases, four key dimensions -- service recipient, target operand resource, service content, and pricing model -- were identified to structure the taxonomy. This taxonomy aided in understanding existing DT-based services and design principles as well as served as a  tool for service providers venturing into developing new DT-enabled services.
In \cite{Pronost2021} classification is exploited to characterize, categorize and distinguish different DTs and related applications.
Whereas in \cite{VanDerValk2020}  the relationship between DTs and simulations, aiming to bridge the gap between their conceptual definitions and real-world applications, were investigated. A taxonomy based on simulation characteristics, analyzing 69 DT applications predominantly in manufacturing has been proposed. The findings reveal a disparity: while simulations are vital to DTs, the two concepts remain distinct, lacking alignment in crucial areas like data connectivity and distribution aspects. This underscores the need to refine definitions for better synergy in analyzing and optimizing systems.

\DT concept named \textit{Universal Digital Twin} is presented in \cite{Akroyd2021}, based on a knowledge graph and ontologies. Data is stored and connected semantically by Uniform Resource Identifiers (URI). Agents are being defined that fulfill tasks on the knowledge graph including data preparation or simulation.
\cite{lyan2021} propose a framework called DATATIME. Models capture at runtime  the state of a socio-technical system according to the dimensions of
both time and space. Space is modeled as a slowly evolving
directed graph with states associated to nodes and edges.

Benaben et. al. present in \cite{Benaben2016} a metaconcept which is defined as a model of the crisis situation itself including partners, objectives, behaviour and context. The technical level includes simulations based on data gatherd from the observed situation and how actors react to it.

The concept of \DTs encompasses a wide range of systems, from small, well-defined machines to large, complex systems of systems. This diversity of scope adds to the complexity of defining what a \DT \cite{Kritzinger2018} is.
Each area, with its own scope and application, has developed its own approach.
While digital twins in production have a strong focus on storing the manufacturing and usage history of products, other applications focus on the lifecycle of products with a strong 3D/CAD aspect from the design phase to production.
For infrastructures and smart cities, the focus is on modeling the functionality and capturing all types of city data. It is an open research task how to design a digital twin for these different areas and first steps towards implementation have already been taken. Crisis management with its big data requirements raises additional areas of research.

\subsection{Previous related work}

\begin{figure}[ht]
\centering
\begin{minipage}{.48\textwidth}
  \centering
  	\includegraphics[width=\columnwidth]{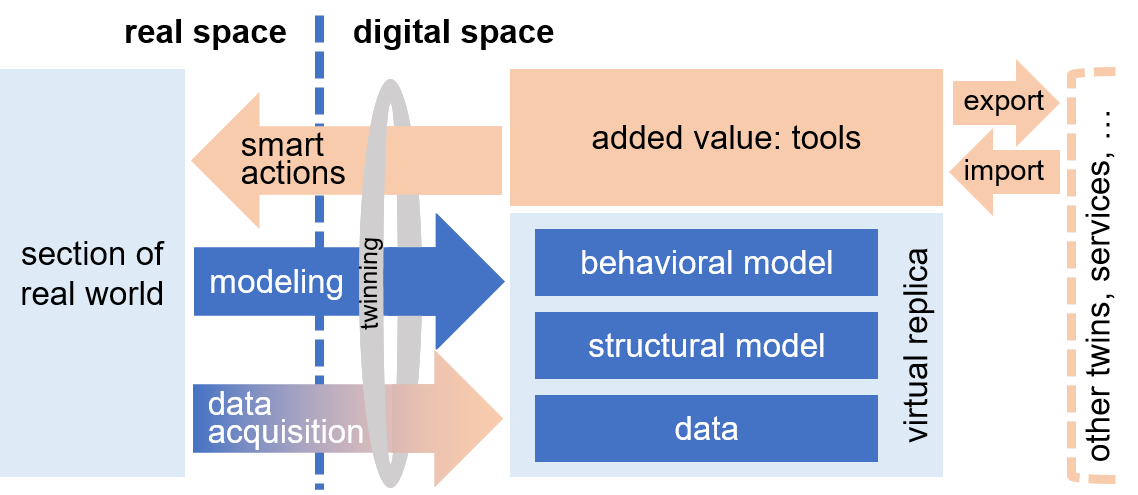}
	\caption{General architecture of a Digital Twin to provide added value \cite{Brucherseifer2021}. The twinning is an essential function of the DT, connecting physical and real space in real-time. The virtual replica consists of models and data.}
    \label{fig:digital_twin_basic}
\end{minipage}
\hfill
\begin{minipage}{.48\textwidth}
   \centering
	\includegraphics[height=0.65\columnwidth]{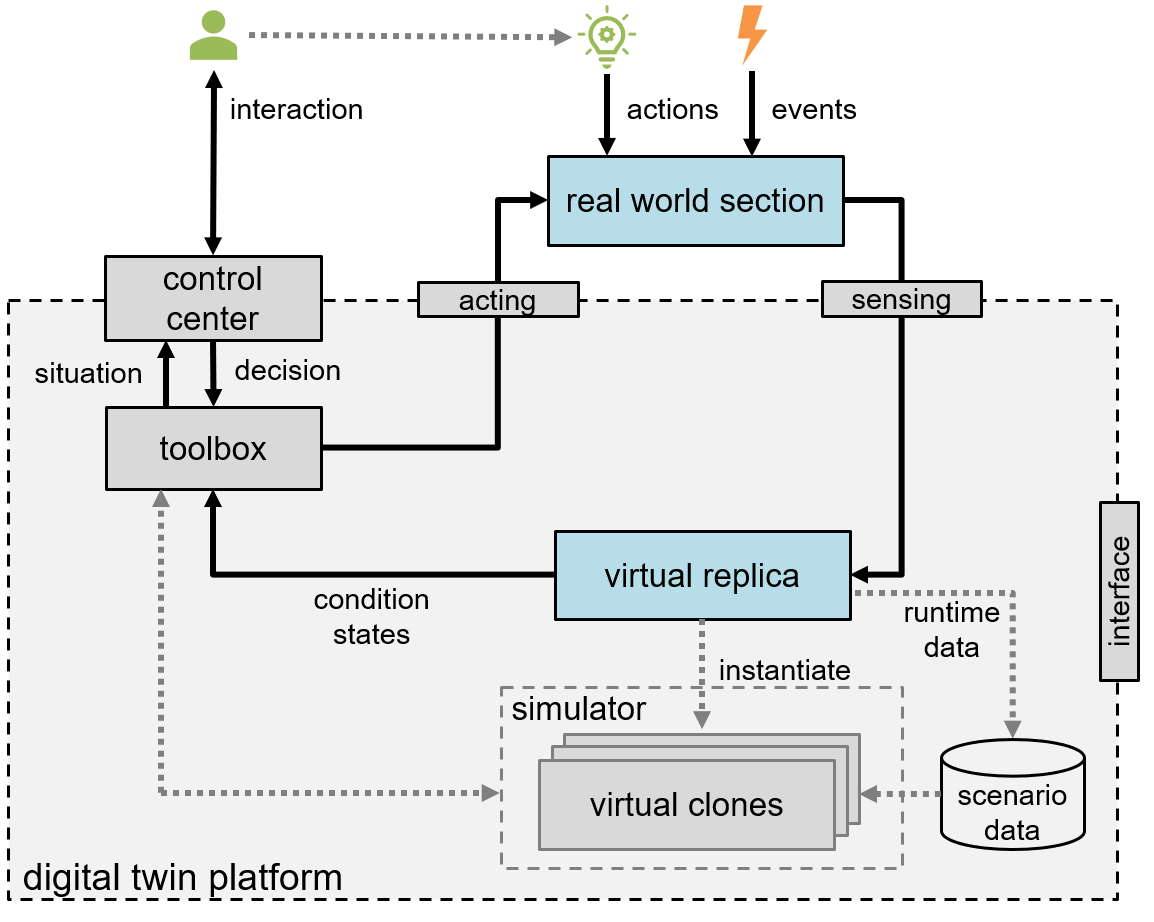}
	\caption{\DT Conceptual Model for the operation of infrastructures to improve resilience in crisis situations, as introduced in \cite{Brucherseifer2021}}
	\label{fig:dt_framework_previous}
\end{minipage}
\end{figure}

In \cite{Brucherseifer2021} we proposed a specific conceptual framework for implementing a \DT to increase the resilience of critical infrastructures in crisis situations, depicted in Figure \ref{fig:dt_framework_previous}. The building blocks and data handling presented in Section \ref{sec:building_blocks} directly inherits from this framework.
The framework is supplied with a study, where we derived implementation requirements from methods of critical infrastructure, crisis management and resilience, compared them with the concept of the digital twin and supplemented them with guiding principles.
The proposed model is suitable for daily use and for crisis situations, with user machine interfaces for situation control and tool interaction. In addition, scenario recording enables retrospective analysis, and simulators based on virtual clones serve as valuable tools for experimentation and optimization and help prepare for rare events and crises. Overall, the application of the digital twin concept to critical infrastructures has been shown to support learning, address challenges in networked systems and systematically improve resilience.
The Digital Twin concept, with its main components meeting the outlined requirements, is presented as a valuable approach to significantly increase the resilience of critical infrastructures. The proposed framework aims to optimize and manage the performance of infrastructures to enhance their resilience.

\section{Methodological Approach and Research Questions}
\label{sec:research_questions}

Crisis management relies on data for multiple parameters supplying information of the current state of the crisis.
In this context a \textit{Data Type} describes a certain type of information e.g. on traffic status. \textit{Data Sources} provide this information and can consist of a
single type of sensor or also a sensor network, that has common characteristics and provides similar information.
Automated collection of data is required for the generation of a situation picture or to keep the virtual replica of a DT on par with the physical world. This requires continuous and reliable input from data sources.
Crisis situations in particular pose a challenge for the provision, as sensors can be destroyed, systems might be operated outside of their design parameters and additional data might be required.
Therefore crisis management systems and specifically DTs needs to be equipped with fallback mechanisms or alarms in case of missing information and thus unreliable work modes.
Replacing or enhancing the data sources for a certain data type by alternative data sources represents a challenge, as properties of each data source may vary from each other even if they belong to the same data type and provide information about the same parameters. Ideally a data source selection procedure should be simple and interpretable for operators with support by the DT itself.

The analysis of the above related works revealed that classification methods and taxonomies for data source management within Digital Twins for disaster management has not been fully explored. Especially the probablity of failure of sources of information causes a severe issue to the applicability of Digital Twins in crisis manamagement.
Hence, as illustrated in the next sections, this paper tackles this gap by (i) employing a methodological approach to identify and comprehend data sources and their interconnections through various attributes, (ii) establishing an assessment function to evaluate the vulnerability during crisis and potential replacements of data sources and related information, and (iii) identifying building blocks of DTs to integrate improved management of data sources. As an overall goal we address simplistic methods as a first step to explore the field.

This leads to the following two questions with respect to data sources (DS) of a common data type:
\begin{itemize}
   \item DS1 (similarity): How similar are the properties of one data source to other available data sources and can they be used as fallback or enhancement?
   \item DS2 (vulnerability): To what extent might the individual sources be affected by the course of a crisis?
\end{itemize}

Taxonomies are a way of organizing and classifying objects to create a structured hierarchy. The objects are grouped according to their characteristics, attributes, and relationships and placed into categories and subcategories. A taxonomy helps  categorizing objects to access and use them easily. In the following section we apply this concept of taxonomy and classification to the domain of data sources in crisis situations.
The attributes of data sources allow to design a similarity measure as a base to identify fallback sources. In case of a sensor loss the alternative source shall contain similar data with similar characterics, yet also with less risk of failure within the given type of crisis. We present the results of this approach in Section \ref{sec:ds_classification}.

The modules designed to answer research questions DS1 and DS2 need to be integrated in the proposed \DT framework. This leads to the following questions with regard to data processing:

\begin{itemize}
	\item P1 (building blocks): Which building blocks of the Digital Twin support data quality?
	\item P2 (data source management): How can incoming data be prepared for further handling in Digital Twins considering the temporal context?
\end{itemize}

In order to answer these questions P1 and P2 we analyze the \DT framework presented in our previous work \citep{Brucherseifer2021} by identifying building blocks for the required functionalty of a crisis management \DT. We design a data source manager to be integrated into the DT.
Thus Section \ref{sec:dt_framework} proposes an approach to integrate sofisticated data source management into the \DT framework.

With this methodological framework, we address the identified research gap while providing practical, applicable solutions to the challenges of data management in crisis situations. The following sections describe in detail the individual components of our approach, the methods, their implementation, and potential impact on disaster management practice.

\section{Data Source Taxonomy and Replacement Mechanism}
\label{sec:ds_classification}

Our approach for improved data source management encompasses three key components designed to enhance the resilience and adaptability of DTs in crisis situations:
The first component is the data source identification and interconnection analysis. In \ref{sec:classification} we propose a classification sheme based on a systematic methodology to identify diverse data sources and comprehend their intricate interconnections. This is achieved through a detailed examination of various attributes, enabling a holistic understanding of the data ecosystem relevant to disaster management. For the second component
we derive a vulnerability assessment and replacement strategy. In \ref{sec:similarity}  we establish a robust assessment function that serves a dual purpose. Firstly, it identifies potential replacements for compromised data sources, ensuring continuity of critical information flows. Secondly, it evaluates the vulnerability of data sources during crisis events. In \ref{sec:filtering} we outline the assessment procedure and in \ref{sec:replacement} a replacement mechanism for data sources. This is the base for a data source manager providing in the digital twin architecture
This third component is the integration into the DT presented in Section \ref{sec:dt_framework}.
In \ref{sec:ex} we apply the proposed method in an example scenario.

\subsection{Taxonomy Scheme}
\label{sec:classification}

\begin{figure}[t]
    \begin{center}
    \includegraphics[width=0.6\columnwidth]{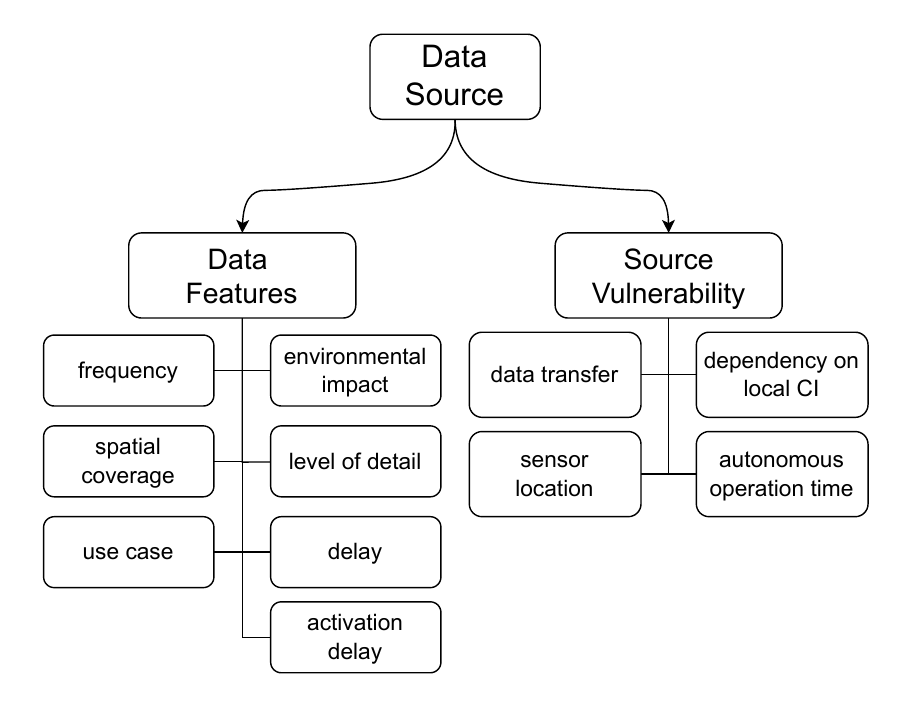}
    \caption{Classification of data source properties. The classification is divided into two main categories, Data Features and Source Vulnerability, each of which contains a number of descriptive characteristics.}
    \label{fig:classification_diagram}
    \end{center}
\end{figure}

As defined in \ref{sec:research_questions}, each Data Source represents processed input data for further usage within the crisis management process or Digital Twin (DT), independent of the actual data sources and sensors used. In our assessment process, this approach allows for handling data sources originating from multiple sensor inputs that are merged during processing, or a single sensor used for multiple data sources.The proposed taxonomy scheme, as illustrated in Figure 2, contains two main categories:
\begin{enumerate}
	\item \textbf{Data Features:} This category addresses the similarity of data characteristics (addressing research question DS1).
	\item \textbf{Source Vulnerability:} This category focuses on characteristics related to the potential impact on performance in case of a crisis event (addressing research question DS2).
\end{enumerate}

Table \ref{tab:attributes} provides a detailed overview of these attributes, including descriptions and examples for each.
Each of these categories contains a selection of attributes identified as having the most significant impact when a fallback needs to be identified.
The vulnerability of the observed parameters depends on the probability of failure of individual data sources. This probability varies for each data source depending on the specific crisis type. For example, in a flood scenario, sensors located in flood-prone areas would have a higher vulnerability compared to those in elevated locations.

\begin{table}[ht]
    \centering
    \scriptsize
    \begin{tabular}{|lll|}
        \hline
         \multicolumn{3}{|c|}{\textbf{Data Features}}\\
         \hline
         \multicolumn{1}{|c}{\textbf {Attribute}} & \multicolumn{1}{c}{\textbf{Description}} & \multicolumn{1}{c|}{\textbf{Example}} \\
         \hline
          environmental impact & possible impact from the (non crisis) environment & weather, time of day\\
          level of detail & detail of the data source & pixel resolution, number of sensors\\
          delay & time to availability for data processing & $x$ seconds \\
          frequency & update interval & $y$ minutes\\
          spatial coverage & spatial coverage & coverage of observed area ($\%$) \\
          activation delay & time until data source is available & $z$ minutes \\
          use case & critical infrastructure monitored with data source & traffic, water supply\\
          \hline \hline
          \multicolumn{3}{|c|}{\textbf{Source Vulnerability}} \\
          \hline
          \multicolumn{1}{|c}{\textbf{Attribute}} & \multicolumn{1}{c}{\textbf{Description}} & \multicolumn{1}{c|}{\textbf{Example}} \\
          \hline
          data transfer & transfer medium to processing & wired, radio, physical \\
          sensor location & placement of sensor - relative to observed area & in situ, remote \\
          dependency on CI & dependency on local critical infrastructure & power supply\\
          autonomous operation time & duration of operation after interruption of supply & $a$ hours \\
         \hline
    \end{tabular}
    \caption{Overview and description of attributes used to assess all sources of information in the context of similarity of features (DS1) and vulnerability during crisis (DS2)}
    \label{tab:attributes}
\end{table}

\subsection{Similarity Function}
\label{sec:similarity}

To evaluate the relationship between different data sources, we propose an assessment function based on a similarity measure. This function calculates the similarity for both Data Feature attributes and Source Vulnerability attributes for all pairs of data sources. The similarity $S$ is defined as the sum of all attribute evaluations $A$ for the data sources $m$ and $n$:

\begin{align}
    S(m,n) =& \sum_{i} w_{i} \cdot A(m,n)_{i} \\
    A(m,n)_{i} =&
    \left\{
      \begin{array}{rl}
    1 & \textrm{for similar characteristics}\\
    0 & \textrm{for similar characteristics with restrictions} \\
    -1 & \textrm{for for different characteristics}
  \end{array}
    \right.
\end{align}

Every attribute A is multiplied by a weighting factor $w_{i}$ to respect the importance of different attributes.

For Data Features, a high similarity value represents a data source with similar characteristics to the reference source. For Source Vulnerability, a high value indicates a similar vulnerability profile. An ideal alternative to the initial data source would have a high similarity scoure in Data Features and a negative similarity score regarding the Source Vulnerability.

\subsection{Assessment Procedure}
\label{sec:filtering}

\begin{figure}[ht]
    \begin{center}
    \includegraphics[height=0.12\columnwidth]{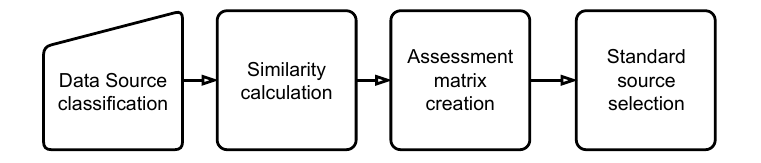}    \caption{Sequence of steps taken to evaluate the attributes of the data sources. }
    \label{fig:assessment_flowchart}
    \end{center}
\end{figure}

The assessment procedure will be performed as a step of preparation upfront and works out the similarities and vulnerabilites of data sources.
Furthermore as a baseline selection the standard sources will be selected for each use case.

The actual assessment procedure consists then of the following steps:
\begin{enumerate}
\item \textbf{Data source categorization:} All available data sources for crisis management are categorized according to the classification scheme.
\item \textbf{Similarity calculation:} The similarity function is applied to all pairs of data sources within each data type or interaction type.
\item \textbf{Assessment matrix creation:} An assessment matrix is created, containing similarity scores for all data source pairs.
\item \textbf{Standard source selection:} The standard source of each information is defined as already implemented in the field or is common practice. The classification results can support the evaluation of sources.
\end{enumerate}

This procedure is performed in advance of any crisis and stored for later use, allowing for quick decision-making during an actual event.

\subsection{Replacement Mechanism during Crisis}
\label{sec:replacement}

\begin{figure}[ht]
    \begin{center}
    \includegraphics[height=0.3\columnwidth]{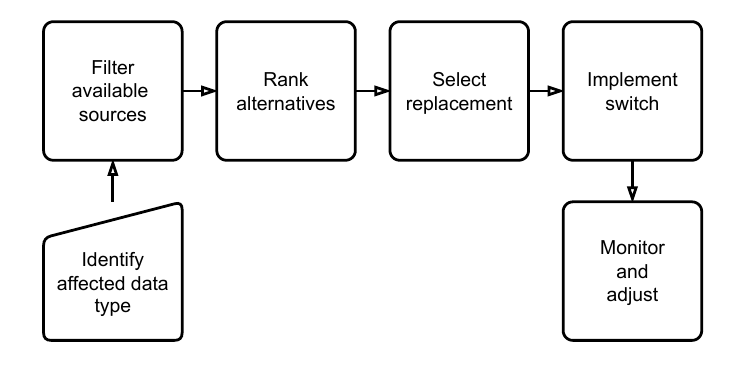}
    \caption{Sequence of steps taken to identify a data source replacement. }
    \label{fig:replacement_flowchart}
    \end{center}
\end{figure}

In the event of a sensor failure or data source loss during a crisis, the following replacement mechanism is activated:
\begin{enumerate}
\item 	\textbf{Identify affected data type:} Determine which data type or interaction type is affected by the loss.
\item	\textbf{Filter available sources:} From the pre-calculated assessment matrix, filter out unavailable or compromised data sources.
\item	\textbf{Rank alternatives:} Rank the remaining alternative data sources based on their similarity scores to the standard source, prioritizing those with high data feature similarity and low vulnerability.
\item	\textbf{Select replacement:} Choose the highest-ranking alternative as the replacement for the lost data source.
\item	\textbf{Implement switch:} Implement the switch to the new data source in the Digital Twin system.
\item	\textbf{Monitor and adjust:} Continuously monitor the performance of the new data source and adjust if necessary.
\end{enumerate}
This mechanism allows for rapid and informed decision-making in selecting alternative data sources during a crisis, ensuring the continued operation of the Digital Twin and supporting effective disaster management.

\section{Applying the Digital Twin Conceptual Framework to Disaster Data}
\label{sec:dt_framework}

In \cite{Brucherseifer2021}, we proposed a Digital Twin Conceptual Framework for the management of resilient infrastructures, as depicted in Figure \ref{fig:dt_framework_previous}.
In the following section, the components of the framework will be identified and their relationship to crisis management tools will be elucidated.  From this set of building blocks and tools, we derive solutions to the research questions P1 and P2, which pertain to the implementation of enhanced and expedient data source management for crisis management.

\subsection{Building Blocks of the Digital Twin Framework}
\label{sec:building_blocks}

\begin{figure}[ht]
\centering
\begin{minipage}{.48\textwidth}
  \centering
	\includegraphics[height=0.65\columnwidth]{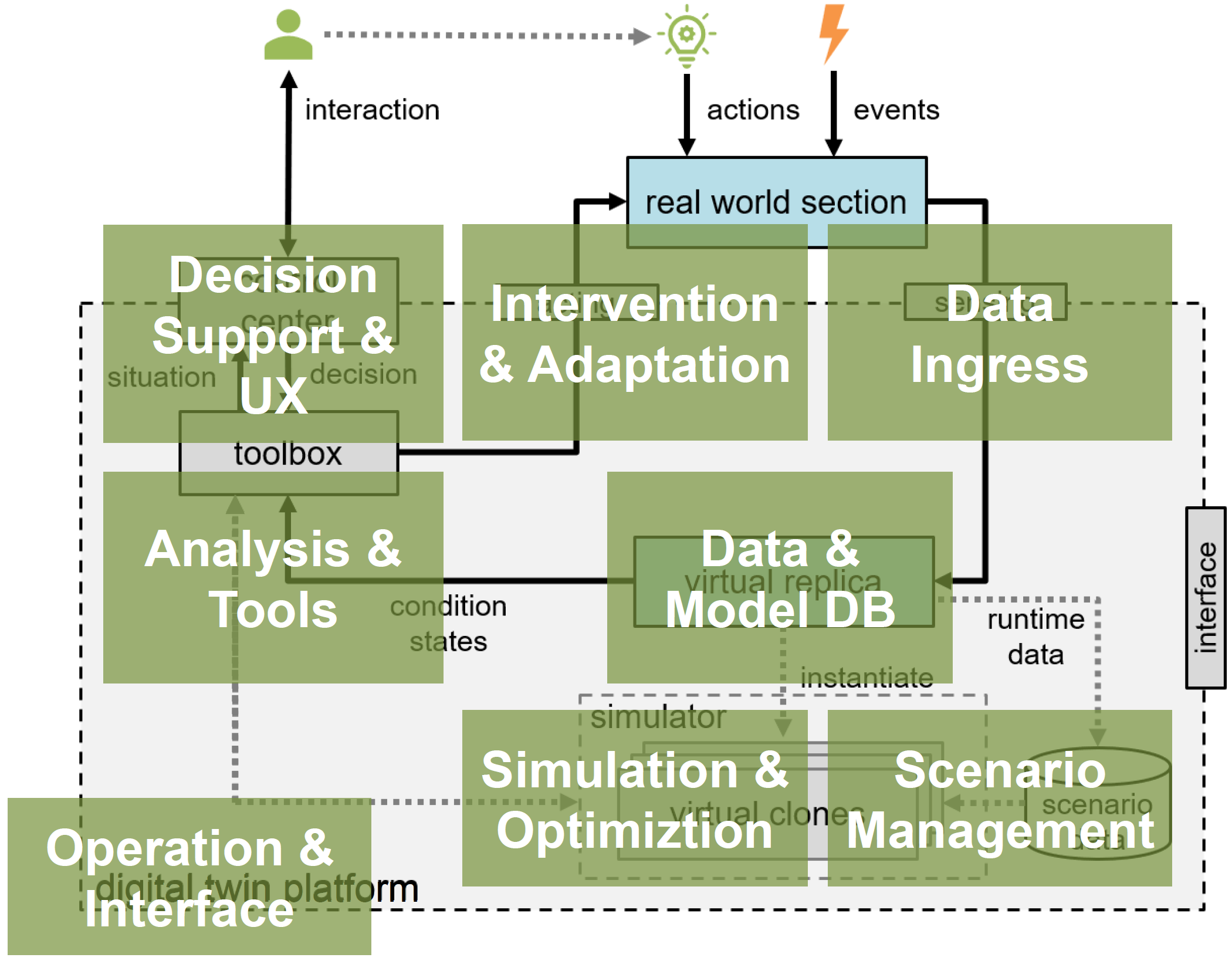}
		\caption{The various bulding blocks contribute to the added value of the \DT Conceptual Model for the operation of infrastructures to improve resilience in crisis situations, as introduced in \cite{Brucherseifer2021}.}
\label{fig:dt_framework}
\end{minipage}
\hfill
\begin{minipage}{.48\textwidth}
   \centering
	\includegraphics[height=0.75\columnwidth]{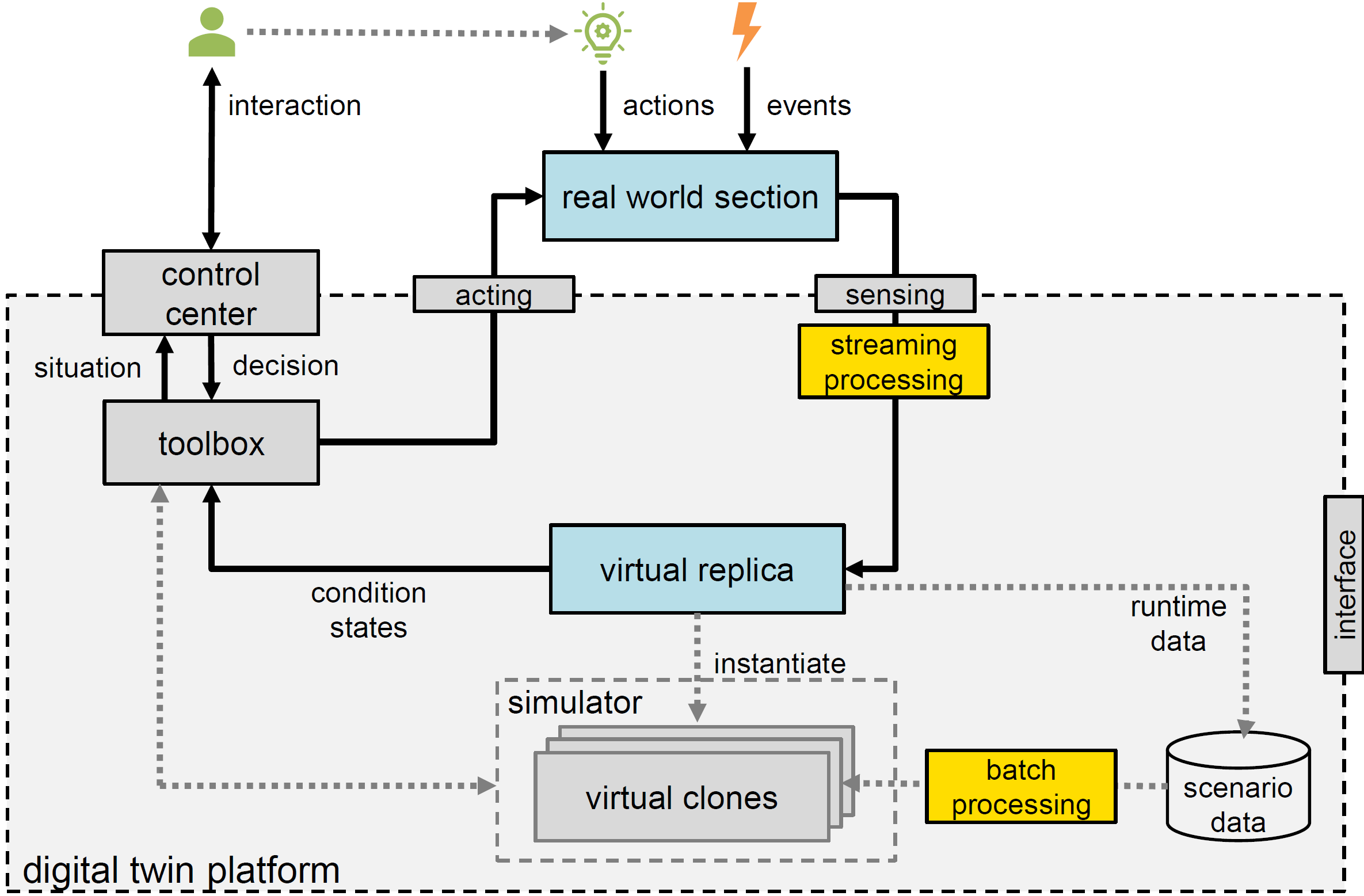}
	\caption{The Digital Twin Conceptual Model enhanced by data processing modules within the Data Ingress and the Scenario Management building blocks.}
	\label{fig:dt_framework_data}
\end{minipage}
\end{figure}

In order to apply the proposed framework to an infrastructure system or urban area, it is first necessary to identify the building blocks that will facilitate this process. The implementation of these building blocks is a prerequisite for the creation of a digital twin, including those related to data sources. The following building blocks were identified:

\textit{Operation \& Interfaces}:
This component provides an operational platform and interfaces to other systems and DTs.
The creation of a DT presents a number of significant challenges, including the integration of diverse data sources into the framework, the provision of interfaces to other systems, the development of tools for the creation and evaluation of new solutions, the addressing of platform scalability and reliability, as well as the standardization and modularization of components.
The formation of a federation of digital twins allows for the mapping of regulatory requirements to infrastructures.

\textit{Data Ingress}:
The Data Ingress component is tasked with the management of all incoming data.
To ensure optimal functionality, the condition of an urban area and its supporting infrastructure is monitored. Examples of data sources include sensors connected to the DT by means of Internet of Things (IoT) mechanisms, information from public and non-public sources such as geographic information system (GIS) data and social media, and the operational status of infrastructure networks (ideally provided by network operators). Infrastructure status may be defined in terms of the availability of essential services, such as power or water supply, and the status of transportation networks. Subsequently, all data undergoes preliminary processing, after which further information is extracted.

\textit{Data \& Model DB}:
The data and model database serves as the virtual replica of the DT. It contains the current status data and all models as a database for other components, thus providing a comprehensive repository of information.
The construction of a virtual replica necessitates the capture of the relevant processes, structure, and behavior of the corresponding real-world area. The current system states and models can be made available via a database that is updated on a regular basis by the data input component and represents a consistent view of the world.
The stored models serve to capture the behavior of the corresponding real section. Models that encompass both normal operations and crisis scenarios, including malfunctioning states, provide the potential for simulations of disruptive occurrences.

\textit{Simulation and Optimization}:
The simulation component enables the replication of the models provided by the virtual replica, thus facilitating the forecasting of potential outcomes, the simulation of alternative scenarios, and the assessment of the impact of interventions. Optimization algorithms identify potential enhancements based on an evaluation function, thereby facilitating the identification of optimal solutions.

\textit{Scenario Management}:
The scenario management component comprises a database that stores historic system states and their respective timelines for retrospective evaluation. This enables the future optimization of processes and enhanced preparedness for crises.
It is crucial to consider scenarios that are pertinent to the monitored critical infrastructure, such as extreme weather events, flooding, or heat.

\textit{Analysis and Tools}:
The component offers a comprehensive array of integrated tools, presented in the form of a toolbox within the user interface.
The objective of the Digital Twin is to enhance resilience capabilities through the utilisation of smart tools, as outlined in Table \ref{tab:resilience_tools}. The activities of these tools are orchestrated over time in order to achieve optimal effectiveness during the various phases of the resilience cycle. It thus follows that the toolbox represents the central component, serving as the nexus for interconnection with all other components.

\textit{Decision Support and UX}:
This component implements methods for decision support and user interaction (UX).
The control center, which serves as the human-machine interface for the DT with regard to stakeholders, provides an intelligent situation overview, offers guidance in navigating situations, and provides access to the toolbox.
Consequently, it assists stakeholders in comprehending the situation and utilizing the requisite tools for resilience activities, while also facilitating remote collaboration. The control center is of vital importance for both routine operations and crisis management.

\textit{Intervention and Adaptation}:
The analysis and plans developed with the support of the DT tools enable stakeholders to implement steps of intervention during a crisis and enhance the adaptation of systems. \\

All components are designed to operate continuously, encompassing the full spectrum of infrastructure operations, including both routine and emergency scenarios. Furthermore, this defines the quantity of data that must be managed by the data ingress component. In a real-world context, such as an urban area or city, it is reasonable to assume that there will be a significant amount of data to be processed.

In order to ensure the implementation and operation of such a \DT platform at all times, as well as during a disaster event, a platform is required to operate the core software services, including data management. It is imperative that the system adheres to established IT standards for secure operation and continuous change. A service-oriented architecture (SOA) is a common approach to ensuring modularity, standardization, and scalability \cite{erl2009}.
In recent years, a multitude of novel big data practices have emerged, including system architectures for big data analytics such as Lambda and Kappa \cite{Marz2015}\cite{Shah2019}. Additionally, a variety of tools and open-source software systems have been developed, including Apache Kafka and Hadoop. Commercial systems, such as Amazon Web Services (AWS) or Microsoft Azure, are available for use. Such systems can be orchestrated to create a crisis management toolkit based on the Digital Twin concept.

\subsection{Dependence of resilient tools on the building blocks}

\begin{table}[ht]
\centering
\scriptsize
\begin{tabular}{|l|l|l|l|}
\hline
\textbf{Activity} & \textbf{Expected Outcome} & \textbf{Tools} & \textbf{Involved Part of DT}\\
\hline
monitor  & situational awareness & anomaly detection & data ingress, analysis\\
 & (continuously) & risk monitoring & data ingress, analysis, decision support, UX \\
&& resilience monitoring & data ingress, analysis, UX \\
&& scenario identification & data ingress, analysis, UX \\
&& scenario forecast & scenario, simulation \\
\hline
respond  & impact mitigation & identification of interventions & scenario, simulation\\
& (during crisis) & intervention forecast & simulation \\
&& decision support intervention & simulation, analysis, decision support, UX \\
&& plan of action & simulation, analysis \\
&& stakeholder alignment & UX \\
\hline
learn & understanding of crisis & replay of scenarios & scenario, simulation, analysis, UX\\
& (during and after crisis) & retrospective analysis & scenario, simulation, analysis, UX \\
&& cause \& reactions & scenario, simulation, analysis, UX \\
&& risk development & scenario, simulation, analysis, UX \\
&& scenario database & scenario \\
\hline
enhance  & refined system & system optimization & scenario, simulation, optimization, analysis  \\
& (after crisis) & decision support, optimization & optimization, analysis, UX \\
&& plan of adaptation & UX \\
&& stakeholder participation & UX, tools, interfaces \\
&& staff training & simulation, UX\\
\hline
\end{tabular}
\caption{Smart tools can provide four types of activities to support infrastructure resilience along the resilience cycle. They concern different parts of the \DT framework. (Abbreviations: ``scenario management = scenario'', ``analysis \& tools = analysis'')}
\label{tab:resilience_tools}
\end{table}

The objective of the DT framework is to enhance the resilience of connected infrastructure systems.
Table \ref{tab:resilience_tools} illustrates the functional activities and anticipated outcomes of the diverse tools provided by the DT framework. The collective functionality of these building blocks equips the DT with resilience capabilities in accordance with the resilience circle. Such activities encompass the monitoring and response to potential disruptions in a given area, the analysis of subsequent events, and the enhancement of infrastructure systems and procedures.

In this framework the Data Ingress component provides the basic data collection for the virtual replica, which in turn enables the digital twin to monitor, analyze and learn from past crises.
Such a system can only function effectively if it is provided with reliable, up-to-date information representing the current state of the infrastructure, both under normal operating conditions and in exceptional system states during crises.
It is essential to consider the challenges of accuracy, uncertainty, and the trade-off between performance and detail of data when developing simulators and tools. The occurrence of disruptive events can result in the interruption of data availability or the introduction of inaccuracies, thereby necessitating the implementation of a resilient design for this component.

Moreover, data is being stored in the Scenario Management building block, which encompasses the database of historical data and system states. It is possible that the data may arrive at runtime with a delay, which may necessitate a rearrangement and correction of the temporal context for subsequent utilisation. The implementation of enhanced data source handling may also result in improvements to data quality and the capacity to simulate and analyse scenarios.

This analysis indicates that the data quality is most influenced by the \textit{Data Ingress} and \textit{Scenario Management} building blocks, thereby answering question P1 of Section \ref{sec:research_questions}.
As numerous tools within the DTs toolbox are contingent upon the availability of high-quality data, the reliability of the DT is enhanced when the data ingress and scenario building blocks are effectively utilized.

\subsection{Integration of Data Classification Method into Digital Twin Framework}
\label{sec:integration}

\begin{figure}[th]
\centering
    \includegraphics[width=0.6\columnwidth]{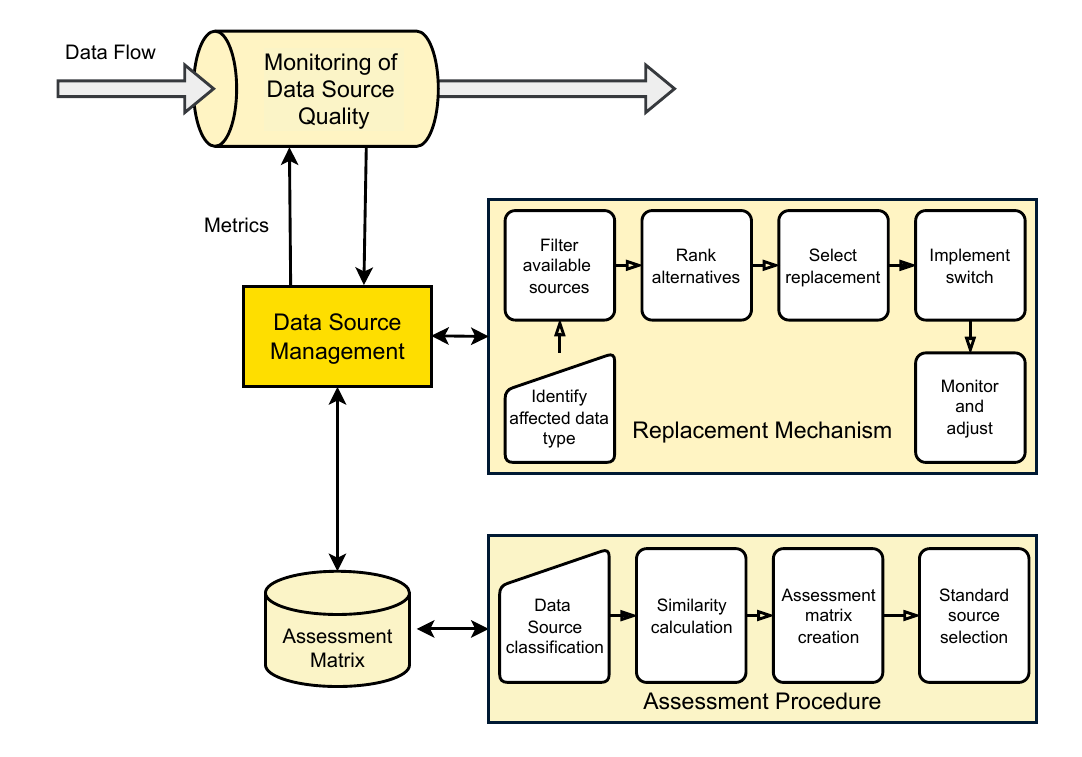}
	\caption{The Digital Twin Conceptual Model enhanced by data handling modules implementing the proposed workflow.}
	\label{fig:dt_framework_source_manager}
\end{figure}

We propose that the new elements, \textit{Streaming Processing} and \textit{Batch Processing} be incorporated into an enhanced DT framework, as illustrated in Figure \ref{fig:dt_framework_data}. These extend the  building blocks of data ingress and scenario management, and perform monitoring of data sources with regard to their status and the plausibility of the incoming data. Stream processing engines are a common component of big data cloud implementations. The monitoring is based on the metrics associated with each data source.
In the event of uncertainty or failure, an alarm can be triggered, prompting the initiation of a replacement mechanism.

In particular, we propose the implementation of a data source management component within the processing units, as illustrated in Figure \ref{fig:dt_framework_source_manager}. The aforementioned component implements the replacement mechanism as previously described in Section \ref{sec:replacement} and is based on the assessment procedure outlined in Section \ref{sec:filtering}. In the event of a change in the evaluation of a data source, the actions outlined in Table 1 can be carried out in accordance with the assessment.
Furthermore, this component facilitates the integration of supplementary data sources that may become available during a crisis.
In the event that new data sources become available during a crisis, such as drone overflight data, an assessment of the additional source needs to be performed, after which integration can take place.

\begin{table}[th]
\center
\footnotesize
\begin{tabular}{|l|l|l|}
\hline
\textbf{Assessment Outcome} & \textbf{Action} & \textbf{Comment}\\ \hline
similar data source already in use & no action & redundancy\\ \hline
similar data source available & activate this data source & fallback\\ \hline
new data source available & integrate this data source & enhanced data\\ \hline
no similar data source available & alarm & no automated action possible \\
\hline
\end{tabular}
\caption{Depending on the result of the assessment different actions can be taken. The classification allows for the identification of data sources that are similar, complementary, or alternative.}
\label{tab:assessment_outcome}
\end{table}

Once the assessment procedure, as delineated in Section \ref{sec:filtering}, has been completed, the algorithm illustrated in Figure \ref{fig:replacement_flowchart} can be integrated into the data ingress building block. The algorithm defines a method for the immediate and continuous processing of incoming data.

The scenario database of the DT framework is designed to store all historical data, including both data received from data sources and generated data. Consequently, it serves as a source of information for batch processing over a specified time interval, representing a crisis event. In the case that data is received from the real world with a delay, it is possible to reorder the data in order to establish the correct chronology. Accordingly, the proposed functional module is incorporated as a preliminary processing stage for the data utilized by the virtual clones for simulation and optimization purposes.

Figure \ref{fig:dt_framework_data} shows an expanded framework that incorporates two novel elements: \textit{streaming processing} and \textit{batch processing}. These elements encompass the specified assessment methodology and replacement mechanism, which are based on the previously proposed classification approach.
The proposed modification of the building blocks, \textit{Data Ingress} and \textit{Scenario Management}, is expected to enhance data source management, thereby improving data quality (question P1).
The algorithms and procedures delineated in Section \ref{sec:ds_classification} can be directly implemented to prepare incoming data with regard to its sources and temporal context (question P2).

\subsection{Discussion}

The approach presented in section \ref{sec:integration} introduced a method of evaluating different data sources.
The entire evaluation process, in addition to the data processing chains for the individual data sources, can be prepared in a pre-crisis state.
The aforementioned steps may be completed in advance, thereby providing a straightforward and efficient method for obtaining results in the event of a crisis.
As the approach is not scenario-specific, it is applicable to all events and depends solely on the available data sources.

The development of robust policies governing data classification and taxonomy within Digital Twins for disaster management is essential. Based on our results we propose the following policies:
\begin{itemize}
    \item [-] Establishing guidelines for standardized classification frameworks could enhance interoperability among systems, ensuring seamless data exchange during crisis.
    \item [-] Additionally, policies that encourage continuous refinement and validation of taxonomies can promote adaptive frameworks, enabling agile responses and improved decision-making in disaster scenarios.
\end{itemize}

On the one hand, the guidelines can then be applied within the streaming processing that synchronize the virtual replica with the real-world area. On the other hand, the scenario data can be pre-processed to control the virtual clones for the simulation of what-if scenarios.

\section{Case study}
\label{sec:ex}

In the following section we examine the possibilities of the data source classification for traffic data in a simplified crisis scenario.
Traffic data plays an important role in crisis management as it helps to organize evacuation efforts and improve the accessibility of critical infrastructure as hospitals.
The scenario consists of a flooding event in an urban area and is described in \ref{sec:ex_scen}.
Flooding events can have a big impact on the transport infrastructure.
As the resulting traffic patterns differ from the pre-crisis traffic patterns, it is important to have an insight into the actual traffic, based on live data.
The selection of traffic data sources taken into account and their assessment is described in \ref{sec:ex_data_features}.
The different steps taken by the digital twin and a crisis manager are described in \ref{sec:ex_workflow}.
\ref{sec:ex_review} concludes the case study with a short review.

\subsection{Scenario description}
\label{sec:ex_scen}

The scenario in \cite{Schneider2024} describes the impact of a 500 year flood of the river Rhine on the critical infrastructure of the city of Cologne, Germany.
The paper focus on the cascading effects of the flooding on the accessibility of hospitals as well as the power supply of the city.
The major factor for the accessibility of hospitals in the paper is the availability of access roads which are unaffected by the flooding.
Additionally, the accessibility could also be affected by congestions on the remaining streets of the city.
Thus, it is necessary to collect live data to provide the current traffic status to the digital twin and a crisis manager.

\subsection{Data Sources and similarity values}
\label{sec:ex_data_features}

\begin{figure}[ht]
\centering
	\includegraphics[height=0.4\columnwidth]{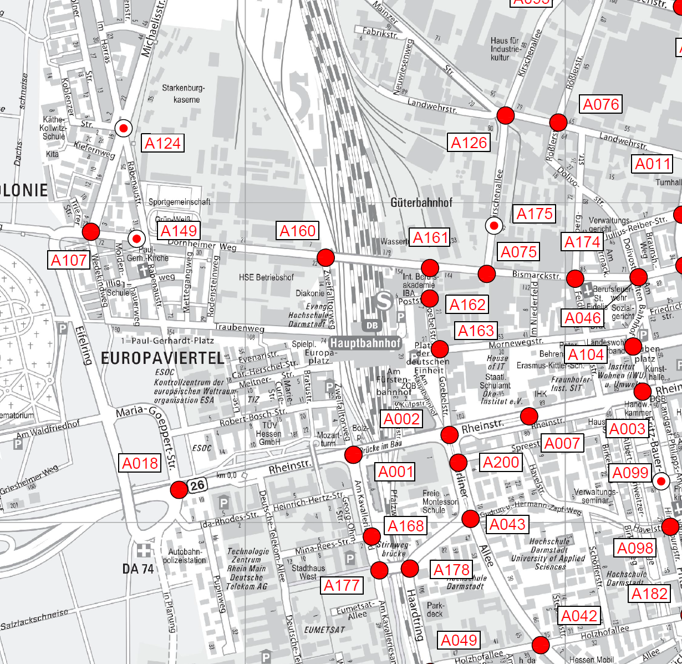}
	\hspace{1cm}
	\includegraphics[height=0.4\columnwidth]{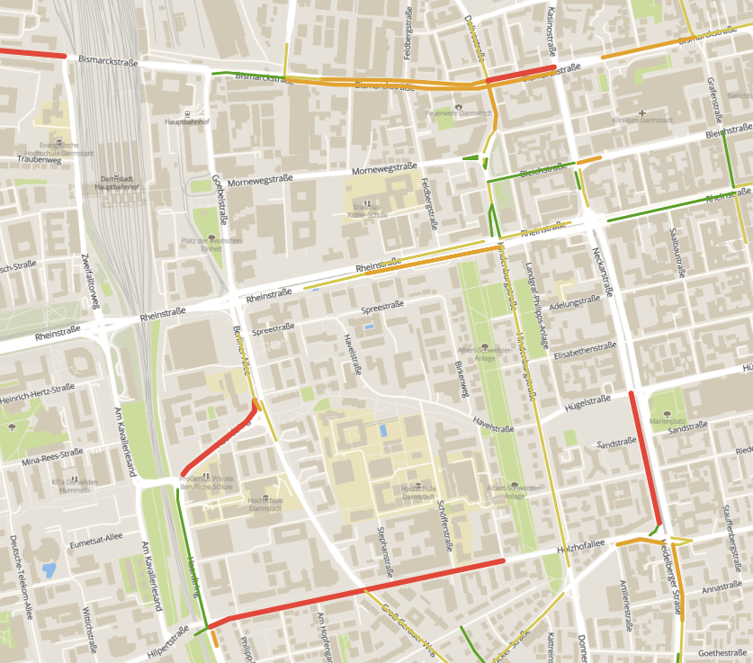}
        \caption[Caption for data sources]{Example of data sources that provide information on traffic and how the data can be visualized: The left graphics shows the locations of traffic lights in the city of Darmstadt, the right one a visual representation of the current traffic data.\footnotemark[1]}
	\label{fig:datasources}
\end{figure}

Traffic density can be calculated from a variety of data sources which provide the number of cars on the road and can be visualised as shown in the right picture of Figure \ref{fig:datasources}. To collect traffic data we selected three different source for this case study:
\begin{itemize}
\item traffic sensor network\\
The traffic sensors are usually located at junctions with traffic lights and installed in the street as inductive loops or located as camera on top of the traffic lights.
They transmit their status to a central traffic server, their location is fixed and distributed over the whole city as shown in Figure \ref{fig:datasources} on the left side.
\item car navigation app data\\
The data from a navigation app transmits the position and speed together with other data from the device, representing the vehicle, to a server for data collection and analysis.
The apps require a cellular data connection to the server for communication.
The number of data points for a particular road depends on the time of day, the use of the road and the distribution of the respective app.
\item remote sensing data from an aerial camera\\
Remote sensing can be carried out using an airborne image detector that flies over the crisis area and sends images to a base station by radio at regular intervals.
At the base station, traffic can be automatically extracted from the images using machine learning algorithms \cite{Stilla2007}\cite{Azevedo2014}.
In addition to traffic detection, the aerial images can also be used for other purposes such as assessing damage to buildings.
\end{itemize}

\footnotetext[1]{Screenshots taken from \url{https://datenplattform.darmstadt.de}, Copyright Wissenschaftsstadt Darmstadt}

The values for each data source (see also table \ref{tab:attributes}) for the traffic data are listed in table \ref{tab:example_input}.
The three main columns represent the different data sources traffic sensors, floating car data and remote sensing data.
Some of the values are publicly available from our cooperation partner, the city of Darmstadt.
If some values were not available we used a best estimate based on literature research or private communication with experts.
With the values for three sources available, the assessment procedure as described in section \ref{sec:filtering} can be carried out.

\begin{table}[ht]
    \centering
    \small
    \begin{tabular}{|p{2,55cm}|p{2,55cm}|p{2,55cm}|p{2,55cm}|}
         \hline
          \multicolumn{1}{|c}{\textbf {traffic}} & \multicolumn{1}{|c}{\textbf {traffic sensors}} & \multicolumn{1}{|c}{\textbf{floating car data}} & \multicolumn{1}{|c|}{\textbf{remote sensing}}  \\

          \multicolumn{1}{|c}{\textbf {detection}} & \multicolumn{1}{|c}{attributes} & \multicolumn{1}{|c}{attributes} & \multicolumn{1}{|c|}{attributes} \\
          \hline \hline
          environmental impact      & none & less data at nighttime & daylight only \\ \hline
          level of detail           & all major street crossings & mainly main roads & $\frac{30 cm}{pixel}$   \\ \hline
          delay                     & 1 sec. & 5 sec. & 1 min.  \\ \hline
          frequency                 & $\frac{1}{s}$  & $\frac{1}{s}$ &  $\frac{1}{h}$ \\ \hline
          spatial coverage          & street crossings with traffic lights & whole city to a different extent & whole city \\ \hline
          activation delay          & none & 1 min. & 20 min. \\ \hline
          \hline
          data transfer             & wired & cellular radio & radio  \\ \hline
          sensor location           & in situ & car & airborne \\ \hline
          dependency on CI          & power & power and network & independent \\ \hline
          autonomous operation time & none & 1h power outage, none network outage & unlimited \\          \hline
    \end{tabular}
    \caption{List of potential data sources for traffic data and the attribute values.}
    \label{tab:example_input}
\end{table}

Table \ref{tab:example_comarisoninput} lists the similarity values between the three sources.
The SUM rows represent the results of the similarity equation, calculated  for the Data Features (top) and the Source Vulnerability (bottom).
Each of the three columns with values represent the comparison of the two Data Source shown in the column heading.
This assessment matrix is calculated well before the crisis and provide the base for the decisions of the replacement procedure of the \DT.

\begin{table}[ht]
    \centering
    \small
    \begin{tabular}{|p{2,55cm}|r|r|r|}
         \hline
          & \multicolumn{1}{|c}{traffic sensors -} & \multicolumn{1}{|c}{traffic sensors -} & \multicolumn{1}{|c|}{floating car data -}  \\

          & \multicolumn{1}{|c}{floating car data} & \multicolumn{1}{|c}{remote sensing} & \multicolumn{1}{|c|}{remote sensing} \\
          \hline \hline
          environmental impact      & 0 & -1 & 0 \\ \hline
          level of detail           & 1 & 1 & 1 \\ \hline
          delay                     & 1 & 0 & 0 \\ \hline
          frequency                 & 1 & -1 & -1 \\ \hline
          spatial coverage          & 1 & 0 & 1 \\ \hline
          activation delay          & 1 & 0 & 1 \\ \hline
          \textbf{SUM}              & \textbf{5} & \textbf{0} & \textbf{1} \\ \hline
          \hline
          data transfer             & -1 & -1 & 0 \\ \hline
          sensor location           & -1 & -1 & -1 \\ \hline
          dependency on CI          & 1 & -1 & -1 \\ \hline
          autonomous operation time & 0 & -1 & -1 \\ \hline
          \textbf{SUM}              & \textbf{-1} & \textbf{-4} & \textbf{-3} \\
         \hline
    \end{tabular}
    \caption{Assessment matrix for three traffic data sources and calculation of the similarity values $S$, for the Data Features (top) and the Source Vulnerability (bottom). }
    \label{tab:example_comarisoninput}
\end{table}

\subsection{Crisis workflow}
\label{sec:ex_workflow}

Prior to the onset of the crisis, the \DT is established and all input data is processed.
Additionally, the assessment matrices are created.
In the case study, the traffic sensors are selected as the baseline Data Source, as they are the typical source for monitoring traffic patterns.
In response to the onset of the crisis, the DT provides traffic information to the control center.
As the water level rises, the power supply to specific areas is either disrupted or switched off as a safety precaution.
This consequently results in the loss of data from the traffic sensors.
The missing Data Source activates the selection procedure from the still available sources.
The procedure aims to have a high similarity value for Data Features as well as a high negative Source Vulnerability.
The result of the aforementioned procedure will prompt the appropriate action, as outlined in \ref{tab:assessment_outcome}.
In case of the flooding the traffic sensors are replaced with data from the floating car data.
Although the transition from one data source to another should be transparent in the visualization at the control center, an indicator is provided to inform users of the change in source.
Consequently, the control center assesses the characteristics of the data sources in \ref{tab:example_input} and determines the differentiating factors between the source and potential alternatives.
As the floating car data source is dependent on the same power supply as the traffic sensors, but has an autonomous operational capacity, an outage in the near future is to be anticipated.
To reduce the activation delay of the next alternative (remote sensing), the control center initiates a process to obtain the aerial images before the anticipated failure of the cellular network.
Upon the availability of remote sensing data, the DT records both incoming data streams of traffic data, while continuing to provide information based on the floating car data.
The continuous availability of data throughout the crisis enables the control center to virtually test solutions to mitigate congestion and maintain hospital accessibility.
Additionally, the control center can adapt the source selection by modifying the weight factors of the assessment matrix, such as increasing the weight factor for autonomous operation time in anticipation of a potential power failure.

\subsection{Scenario review}
\label{sec:ex_review}

In the case study we have shown how the interaction of different data sources with a \DT can support a crisis manager in a flooding scenario with power outages.
While the case study shows only a very small part of the information incoming to a control center, the simplified logic of the data processing helps the crisis manager to understand the behavior of the \DT and take the right decisions.
Also it should be noted that this kind of information will be created and presented to the control center for each data source and even with proper visualization will create a demanding environment for the crisis manager.

\section{Conclusion}
\label{sec:conclusion}

In this paper we propose a comprehensive approach to data classification and management within the context of Digital Twins for disaster management as well as a data source management component within the DT. Central to this approach is the development of a taxonomy for categorizing data sources. In this context, a taxonomy refers to a systematic classification scheme that organizes data sources into categories based on shared characteristics or attributes. The proposed taxonomy considers two main aspects of data sources:
The first focuses on the \textit{Data Feature}. These attributes describe the characteristics of the data itself, such as environmental impact, level of detail, delay, frequency, spatial coverage, and activation delay. The second aspect is the \textit{Source Vulnerability}. These attributes assess the susceptibility of the data source to failure or disruption during a crisis, including factors like data transfer method, sensor location, dependency on critical infrastructure, and autonomous operation time.

In this framework, a \textit{data type} is defined as a specific category of information relevant to disaster management. Examples of data types include traffic data, weather data, and infrastructure status data. Each data type may have multiple associated data sources, which are the specific systems, sensors, or methods used to collect and provide that type of data. To illustrate, traffic data, a specific data type, may be acquired from inductive loop sensors, floating car data, or aerial imagery, which collectively constitute the data sources.

The paper introduces a similarity function to quantitatively assess the relationship between different data sources within the same data type. This function enables the identification of potential replacement sources when primary data inputs are compromised. By evaluating both the similarity of data features and the differences in vulnerability profiles, the proposed approach facilitates intelligent selection of alternative data sources that can maintain the DT's functionality while potentially offering improved resilience to ongoing crisis conditions.

Furthermore, this research outlines a strategy for integrating the proposed data classification and replacement mechanism into existing Digital Twin frameworks. We worked out the building blocks of a \DT, their relation to the intended resilience activities and their reliability on data. Focusing on the \textit{Data Ingress} and \textit{Scenario Management} components of the DT architecture, the paper describes how streaming and batch processing modules can be enhanced by a \textit{data source manager} to incorporate real-time monitoring of data source quality and automated replacement procedures. This component implements the assessment and replacement mechanisms with several potential resulting actions.

The proposal has been demonstrated with traffic data sources.
The results show that the similarity function provides useful insights that help in the assessment of different data sources.
The evaluation allows the selection of data sources with similar performance characteristics and different vulnerability and thus represents an important element for the operation of DT in different crisis situations.

Our work identified data classification as an essential aspect for disaster preparedness and response infrastructure. The ability to identify critical, interchangeable, or complementary elements is crucial for strengthening this infrastructure through the digital twin approach. In crisis situations, it is essential to understand the details, especially when data or information may not be fully accessible. While some data may be irreplaceable, others can be replaced or integrated.   This ability to differentiate is crucial in disaster management for implementing and applying theoretical frameworks related to the design and realization of digital twins. Data classification is the foundation of this system, allowing for the creation of digital replicas that aid in simulating, analyzing, and managing crises effectively.

Future work will aim to expand the proposed classification model by incorporating additional categories or more detailed classifications as required.
This may include the integration of novel data types and technologies to enhance the model's comprehensiveness and adaptability.
The utilization of ontologies, as well as existing data models such as the FIWARE Smart Data Model \citep{Jara2018}, may offer supplementary attributes and the potential to automatically create classifications for data sources.

The current data processing and data selection process is based on a simplistic approach that is comprehensive and adaptable.
These two features are crucial for an effective human-in-the-loop control system and need to be considered when implementing improvements to the data processing.
While deep learning algorithms are known for their lack of transparency, alternative approaches, such as data fusion, could be deployed to supersede the single-source methodology.
For example, data analysis based on Bayes theorem
represents a well-established method with traceable results.
As illustrated in \cite{Schneider2023}, a Bayesian network can support decision-making processes. However, even in a limited case study, its complexity is considerable. Therefore, the implementation of such an approach or any other improvement to data processing requires a balanced approach between optimizing the use of available data and providing an intuitive representation of the decision-making process of the DT to the control center.

\section*{Declaration of AI and AI-assisted technologies in the writing process}

During the preparation of this work the authors used DeepL in order to improve readability and language of the text. After using this tool/service, the authors reviewed and edited the content as needed and take full responsibility for the content of the publication.

\bibliographystyle{plainnat}
\bibliography{literature}

\end{document}